\documentclass[reprint,
 amsmath,amssymb,
 aps,
]{revtex4-2}

\usepackage{graphicx}
\usepackage{dcolumn}
\usepackage{bm}
\usepackage{xcolor}
\usepackage{mathptmx}
\usepackage{graphicx}
\usepackage{dcolumn}
 
\usepackage{bm}
\usepackage{amssymb}
\usepackage{comment}
\usepackage{hyperref}
\usepackage{physics}

\begin{document}

\title{Selective avoidance of multiple line-of-sight obstacles at 130~m using locally bending space-time wave packets}

\author{Layton A. Hall$^{1,2}$}

\author{Murat Yessenov$^{1,3}$}%

\author{Isabelle Lebron$^{1}$}%

\author{Ayman F. Abouraddy$^{1}$}
\email{raddy@creol.ucf.edu}
\affiliation{$^{1}$CREOL, The College of Optics \& Photonics, University of Central Florida, Orlando, FL 32816, USA}%
\affiliation{$^{2}$Materials Physics and Applications - Quantum Division, Los Alamos National Laboratory, Los Alamos, NM 87545, USA}
\affiliation{$^{3}$Harvard John A. Paulson School of Engineering and Applied Sciences, Harvard University, Cambridge, MA, USA}


\begin{abstract}
Self-accelerating optical beams follow curved trajectories rather than propagating rectilinearly, which raises the prospect for avoiding line-of-sight (LoS) obstacles blocking the beam path. However, if a conventional laser beam is intercepted by an obstacle en route to an intended LoS target, then replacing this beam with a bending beam is not adequate: the obstacle is avoided but the target cannot be concomitantly reached. Rather, a laser beam that \textit{locally} bends around the obstacle -- before continuing along its rectilinear path -- is needed. Here we show that engineering the spatiotemporal spectrum of a pulsed beam yields a space-time wave packet whose propagation dynamics can be tuned at will to locally bend around one or multiple obstacles, thereby avoiding them, to selectively reach a designated target. We carry out our experiments over distances extending to 130~m from the source, carried out in an open-air environment. In one scenario, the beam is incident on a target placed between two LoS obstacles, one preceding it and one following it -- both of which are avoided. In a second scenario, the locally bending beam avoids one and then two LoS obstacles preceding the intended target. These results may contribute to applications requiring selective incidence on targets in remote sensing, stand-off detection, directed energy, for optical and radio-frequency communications in presence of LoS obstacles, and for selectively delivered radiation therapies.
\end{abstract}


\maketitle

\section{Introduction}

\begin{figure*}[t!]
\centering
\includegraphics[width=18.4cm]{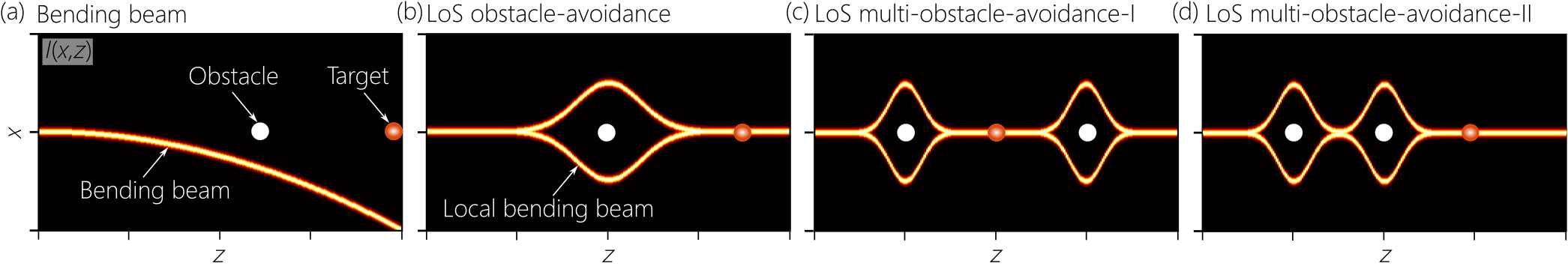}
\caption{(a) Idealized intensity profile $I(x,z)$ for a \textit{bending beam} whose peak follows a power-law trajectory (e.g., an Airy beam \cite{Siviloglou07OL} or a bending STWP \cite{Hall25OLbending}). Such a beam can avoid an LoS obstacle (depicted as a white circle), but cannot concomitantly reach an LoS target (depicted as a red circle) downstream beyond the obstacle. (b) A \textit{locally bending beam} can avoid the LoS obstacle \textit{and} reach the LoS target beyond it. (c,d) Multi-obstacle-avoidance utilizing locally bending beams. (c) In Scenario-I, the beam impinges on a mid-range target while avoiding obstacles before and after it. (d) In Scenario-II, the beam impinges on a target at the end of the range while avoiding two obstacles preceding it.}
\label{Fig:Concept}
\end{figure*}

Self-accelerating optical beams \cite{Efremidis19Optica} that travel along curved trajectories [Fig.~\ref{Fig:Concept}(a)], rather than the usual rectilinear propagation \cite{SalehBook07}, suggest a pathway to avoid obstacles lying in the line-of-sight (LoS) en route to an intended remote target \cite{Polynkin09Science,Clerici15SciAdv}. Rather than follow a fixed bending trajectory, this task requires a beam that bends \textit{locally} around the object before continuing its path to the designated target [Fig.~\ref{Fig:Concept}(b)]. Indeed, selectively avoiding multiple LoS objects is the ultimate test to verify beam-bending functionality. Applications of such locally bending beams include selective targeting in remote sensing, stand-off detection, and directed energy. Another potential application is in laser radiation therapies where it may be necessary to avoid intervening sensitive tissue or organs before the beam reaches the intended target. Moreover, in optical or high-radio-frequency wireless communications, a locally bending beam is required to maintain connectivity in presence of an intervening obstacle in the LoS between the transmitter and receiver \cite{Hu24NC,Guerboukha24CE,Chen25NC}.

To date, optical beam-bending has been verified experimentally using Airy beams \cite{Siviloglou07OL,Siviloglou07PRL} and related structured beams \cite{Bandres08OL,Davis09OE,Morris09OE,Morris10JO,Chremmos12OL,Kaminer12PRL,Zhao13OL,Bekenstein14PRX,BarDavid16SR}. These are usually monochromatic optical beams having the following characteristics: (1) their transverse intensity profile is invariant with propagation (diffraction-free); (2) their transverse profile is spatially asymmetric, and the asymmetry direction determines the orientation of the curved trajectory; (3) the transverse profile features a spatial oscillatory structure whose spatial scale determines the self-acceleration or bending rate \cite{Efremidis19Optica}; and (4) the beam is translated laterally with propagation, usually following a parabolic trajectory \cite{Siviloglou08OL}. 

Most recently, the class of spatiotemporally structured optical fields \cite{Shen23JO,Abouraddy25OPN} known as `space-time wave packets' (STWPs) \cite{Kondakci16OE,Parker16OE,Kondakci17NP,Yessenov22NC}, in which a tight association is introduced between the spatial and temporal frequencies underpinning the field, has been exploited to produce bending beams \cite{Hall25OLbending}. The distinctive features of STWPs verified over the past decade (e.g., propagation-invariance in linear media \cite{Bhaduri18OL,He22LPR,Hall23LPR} at a tunable group velocity \cite{Efremidis17OL,Porras17OL,Wong17ACSP2,Kondakci19NC}, anomalous refraction \cite{Bhaduri20NP}, self-healing \cite{Kondakci18OL}, and axial acceleration \cite{Yessenov20PRLAccel,Li20CP,Li20SR,Hall22OLAccel}, among other features \cite{Yessenov22AOP}) have all accompanied \textit{rectilinear} propagation. We have demonstrated \cite{Hall25OLbending} that engineering the space-time coupling introduced into a generic pulsed beam yields `bending STWPs' (proposed theoretically in \cite{Liang23OL,Wang25OL}), which have the following characteristics: (1) the transverse spatial profiles of the time-averaged intensity profile is invariant with propagation (diffraction-free); (2) they have spatially \textit{symmetric} transverse profiles; (3) the transverse profile features a spatial oscillatory structure, but its spatial scale does \textit{not} determine the self-acceleration or bending rate; and (4) the beam is translated laterally with propagation following a power-law trajectory $x_{\mathrm{o}}(z)=x_{1}(\tfrac{z}{z_{1}})^{\gamma}$, where $x_{\mathrm{o}}$ is the position of the beam peak intensity along the transverse coordinate $x$, $z$ is the axial coordinate, $x_{\mathrm{o}}(0)=0$, the beam shifts laterally to $x_{1}$ when $z=z_{1}$, and $\gamma\geq0$ is an arbitrary positive number, whether integer, fractional, or real \cite{Hall25OLbending}.

However, selectively avoiding a LoS obstacle [Fig.~\ref{Fig:Concept}(b)] or multiple obstacles [Fig.~\ref{Fig:Concept}(c,d)] to impinge upon a LoS target requires a beam that bends \textit{locally} around the obstacles rather than follow a fixed bending trajectory [see Fig.~\ref{Fig:Concept}(a)]. To date, this obstacle-avoidance configuration has not been verified to the best of our knowledge with Airy beams. Furthermore, large-scale measurements in open-air environments that involve multiple obstacles are needed to complement laboratory-scale experiments performed in a well-controlled environment.

Here we show that phase-only, spatiotemporal spectral modulation of a generic pulsed laser beam can yield an STWP whose propagation dynamics is sculpted by introducing complex local beam-bending that allows for selective avoidance of a single or multiple LoS obstacles while maintaining incidence on an intended target. Obstacle-avoidance is demonstrated here in an open-air laser-testing facility ($\sim130$~m of axial propagation) using a pulsed laser at a wavelength $\approx1$~$\mu$m. In one scenario, we sculpt the STWP trajectory to selectively impinge on a target situated between two obstacles (one preceding it and one following it) that we avoid. In this scenario, two nulls are introduced into the intensity profile of the beam along the LoS before and after the intended target at the obstacle locations. In the second scenario, the STWP impinges on a target at the end of the range, and one or two intervening obstacles are to be avoided. Here, two nulls are introduced into the beam path, both preceding the target. The intended targets and obstacles are located at positions extending from 20~m to 130~m from the experimental setup. The beam synthesis follows an algorithmic procedure sufficiently versatile to accommodate a wide variety of obstacle-avoidance scenarios. These results pave the way to exploiting locally bending laser beams in applications ranging from optical and radio-frequency communications to laser radiation therapy, and from remote sensing to directed energy.

\section{Concept of local-bending for line-of-sight target-avoidance}

\subsection{Propagation-invariant STWPs}

We first briefly describe propagation-invariant STWPs that serve as a baseline for their locally bending counterparts. We write the field for a pulsed beam, $E(x,z;t)=e^{i(k_{\mathrm{o}}z-\omega_{\mathrm{o}}t)}\psi(x,z;t)$, in terms of a carrier and slowly varying spatiotemporal envelope expressed in terms of an angular spectrum:
\begin{equation}
\psi(x,z;t)=\iint\!dk_{x}d\Omega\;\widetilde{\psi}(k_{x},\Omega)e^{i\{k_{x}x+(k_{z}-k_{\mathrm{o}})z-\Omega t\}},
\end{equation}
where $\widetilde{\psi}(k_{x},\Omega)$ is the spatiotemporal spectrum, $\omega$ is the temporal frequency, $\Omega=\omega-\omega_{\mathrm{o}}$, $\omega_{\mathrm{o}}$ is a fixed temporal frequency, $k_{\mathrm{o}}=\omega_{\mathrm{o}}/c$, $c$ is the speed of light in vacuum, and $k_{x}$ and $k_{z}$ are the transverse and axial wave numbers, respectively, with $k_{x}^{2}+k_{z}^{2}=(\tfrac{\omega}{c})^{2}$. The spatiotemporal spectrum $\widetilde{\psi}(k_{x},\Omega)$ of a \textit{conventional} pulsed beam is a 2D domain in $(k_{x},\Omega)$-space, having spatial and temporal bandwidths $\Delta k_{x}$ and $\Delta\omega$, respectively, which are -- in principle -- independent of each other. The hallmark of a propagation-invariant STWP [Fig.~\ref{Fig:Theory}(a)] that travels rigidly in free space is a linear constraint on the temporal frequency $\omega$ and the axial wave number $k_{z}$, $\Omega=(k_{z}-k_{\mathrm{o}})c\tan\theta$, where we refer to $\theta$ as the spectral tilt angle \cite{Kondakci19NC}. Consequently, their spatiotemporal spectra are restricted to 1D trajectories in $(k_{x},\Omega)$-space, $\widetilde{\psi}(k_{x},\Omega)\rightarrow\widetilde{\psi}(k_{x})\delta(\Omega-\Omega(k_{x}))$, such that each spatial frequency $k_{x}$ is associated with a single temporal frequency $\Omega(k_{x})$ [Fig.~\ref{Fig:Theory}(b)]:
\begin{equation}\label{eq:Parabola}
\frac{\Omega(k_{x})}{\omega_{\mathrm{o}}}\approx\frac{k_{x}^{2}}{2k_{\mathrm{o}}^{2}(1-\cot\theta)},
\end{equation}
whereupon the spatial and temporal bandwidths are no longer independent of each other \cite{Kondakci17NP}. Crucially, there is a one-to-one relationship between $\omega$ and $|k_{x}|$, which is symmetrically parabolic around $k_{x}=0$ [Fig.~\ref{Fig:Theory}(b)]; note that the spectral phase is flat. This results in a propagation-invariant STWP traveling along the $z$-axis, whose time-averaged intensity profile $I(x,z)=\int\!dt\;|\psi(x,z;t)|^{2}$ is: 
\begin{equation}\label{eq:Intensity}
I(x,z)=\int\!dk_{x}|\widetilde{\psi}(k_{x})|^{2}+\int\!dk_{x}\widetilde{\psi}(k_{x})\widetilde{\psi}^{*}(-k_{x})e^{i2k_{x}x},
\end{equation}
which is diffraction-free $I(x,z)=I(x,0)$, and takes the form of a central spatial feature of width $\Delta x\sim\tfrac{\pi}{\Delta k_{x}}$ (second term in the Eq.~\ref{eq:Intensity}) atop a pedestal whose width is determined by the system aperture (first term in the Eq.~\ref{eq:Intensity}) \cite{Kondakci17NP}.  

\subsection{Bending STWPs}

The parabolic relationship in Eq.~\ref{eq:Parabola} between $k_{x}$ and $\omega$ for a \textit{propagation-invariant} STWP is the baseline spatiotemporal spectrum for \textit{bending} STWPs. The strategy for creating bending STWPs can be appreciated by first examining a \textit{tilted} diffraction-free STWP that travels rectilinearly at an angle $\varphi_{\mathrm{o}}$ with the $z$-axis, whose intensity-peak trajectory is $x_{\mathrm{o}}(z)=z\tan\varphi_{\mathrm{o}}$ \cite{Hall25OLbending}. This field structure results from subjecting a propagation-invariant STWP to the geometric transformation $z'=z\cos\varphi_{\mathrm{o}}-x\sin\varphi_{\mathrm{o}}$ and $x'=z\sin\varphi_{\mathrm{o}}+x\cos\varphi_{\mathrm{o}}$  in physical space; here $(x',z')$ is the new coordinate system for the tilted STWP. Accompanying this transformation in physical space is a corresponding rotation in Fourier space: $k_{z}'=k_{z}\cos\varphi_{\mathrm{o}}-k_{x}\sin\varphi_{\mathrm{o}}$ and $k_{x}'=k_{z}\sin\varphi_{\mathrm{o}}+k_{x}\cos\varphi_{\mathrm{o}}$. Because the spectral support for the propagation-invariant STWP is confined to a plane $\Omega=(k_{z}-k_{\mathrm{o}})c\tan\theta$ in $(k_{x},k_{z},\tfrac{\omega}{c})$-space, which is parallel to the $k_{x}$-axis and is tilted an angle $\theta$ with the $k_{z}$-axis \cite{Yessenov19PRA}, the spectral support for the tilted STWP remains confined to a plane, but this plane is rotated by $\varphi_{\mathrm{o}}$ around the $\tfrac{\omega}{c}$-axis. Consequently, the spectral projection onto the $(k_{x},\tfrac{\omega}{c})$-plane is still a parabola, but its center is displaced from $k_{x}=0$ \cite{Hall25OLbending}.

\begin{figure}[t!]
\centering
\includegraphics[width=8.8cm]{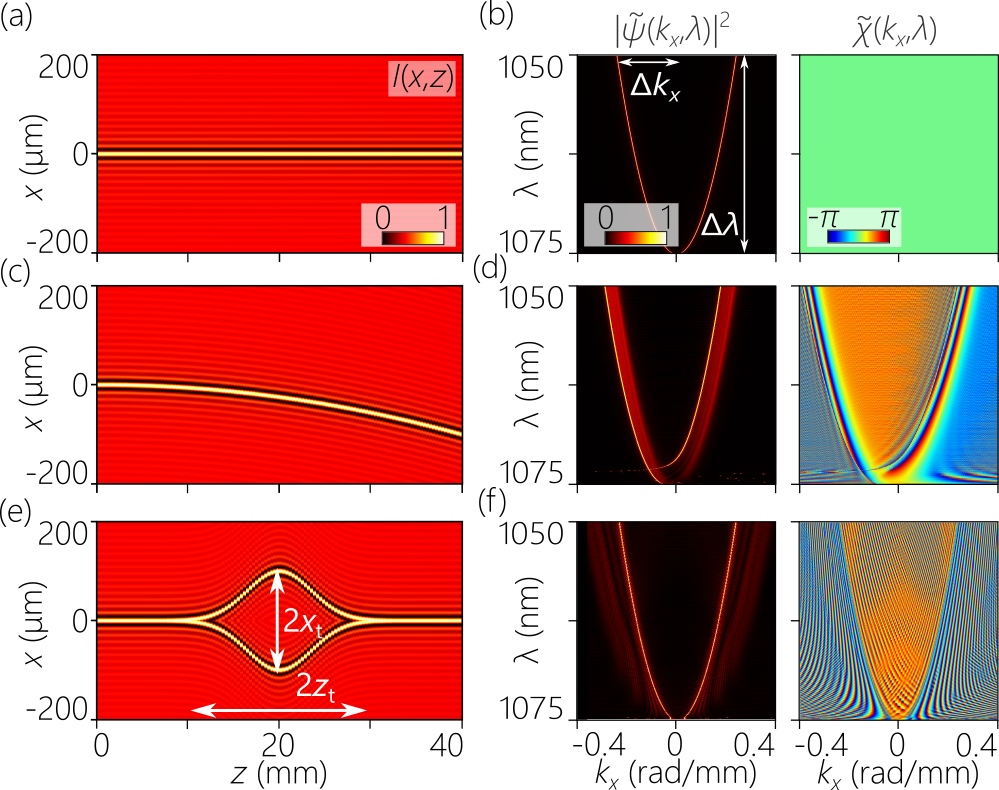}
\caption{(a,b) A propagation-invariant STWP; $\theta=46^{\circ}$ and $\Delta\lambda=25$~nm. (a) The time-averaged intensity $I(x,z)$ and (b) the magnitude-squared $|\widetilde{\psi}(k_{x},\lambda)|^{2}$ and phase $\widetilde{\chi}(k_{x},\lambda)$ of the complex spatiotemporal spectrum, $\widetilde{\psi}(k_{x},\lambda)\!=\!|\widetilde{\psi}(k_{x},\lambda)|e^{i\widetilde{\chi}(k_{x},\lambda)}$. (c,d) Same as (a,b) for a \textit{bending} STWP; $\theta=46^{\circ}$, $\Delta\lambda=25$~nm, following a parabolic trajectory $x_{\mathrm{o}}(z)=x_{1}(\tfrac{z}{z_{1}})^{2}$, $x_{1}=100$~$\mu$m, and $z_{1}=40$~mm. (e,f) Same as (a,b) for a \textit{locally bending} STWP with $x_{\mathrm{o}}(z)=\pm x_{\mathrm{t}}\exp\left\{-(z-z_{1})^{2}/(2z_{\mathrm{t}}^{2})\right\}$, with $x_{\mathrm{t}}=100$~$\mu$m, $z_{1}=20$~mm, and  $z_{\mathrm{t}}=10$~mm.}
\label{Fig:Theory}
\end{figure}

A bending STWP can be viewed as a generalization of a tilted STWP, in which the fixed tilt angle $\varphi_{\mathrm{o}}$ is replaced by an axially varying tilt angle $\varphi(z)$, where $\tan\varphi(z)=\tfrac{x_{\mathrm{o}}(z)}{z}$ [Fig.~\ref{Fig:Theory}(c)]. Heuristically, the rotation in $(k_{x},k_{z},\tfrac{\omega}{c})$-space must also be $z$-dependent. The resulting spectral projections onto the $(k_{x},\tfrac{\omega}{c})$-plane is no longer one-to-one between $|k_{x}|$ and $\omega$, nor symmetric. Rather, each frequency $\omega$ is now associated with a small but finite spatial bandwidth, and the spectral phase is no longer flat \cite{Hall25OLbending}; see Fig.~\ref{Fig:Theory}(d). Similarly to the propagation-invariant STWP [Fig.~\ref{Fig:Theory}(a)], the time-averaged intensity of the bending STWP [Fig.~\ref{Fig:Theory}(c)] is formed of a background term on top of which is a spatially localized transverse feature that travels, however, along a curved trajectory rather than rectilinearly.

\subsection{Locally bending STWPs}

Here we aim at producing STWPs with more complex trajectories that are not described by a power law. For LoS single-obstacle-avoidance, we aim at a trajectory that takes the form:
\begin{equation}
x_{\mathrm{o}}(z)=\pm x_{\mathrm{t}}\exp\left\{-\frac{(z-z_{1})^{2}}{2z_{\mathrm{t}}^{2}}\right\},
\end{equation}
where $z_{1}$ is the axial plane at which the obstacle is located, $z_{\mathrm{t}}$ is the axial length of the null along the optical axis centered at the plane $z=z_{1}$, and $2x_{\mathrm{t}}$ is the transverse width of the null at that plane [Fig.~\ref{Fig:Theory}(e)]. We refer to such a beam as a `locally bending STWP' rather than simply a `bending STWP'. Such a beam is the result of a complex modification of the spatiotemporal spectrum [Fig.~\ref{Fig:Theory}(f)]. Note that we initially expect that the background term will persist in the center of the local bend. We will see below that this is not always the case, and that a null develops in the center of the bend under certain conditions. To understand the preparation of a locally bending STWP, we first describe the synthesis methodology.

\begin{figure}[t!]
\centering
\includegraphics[width=8.6cm]{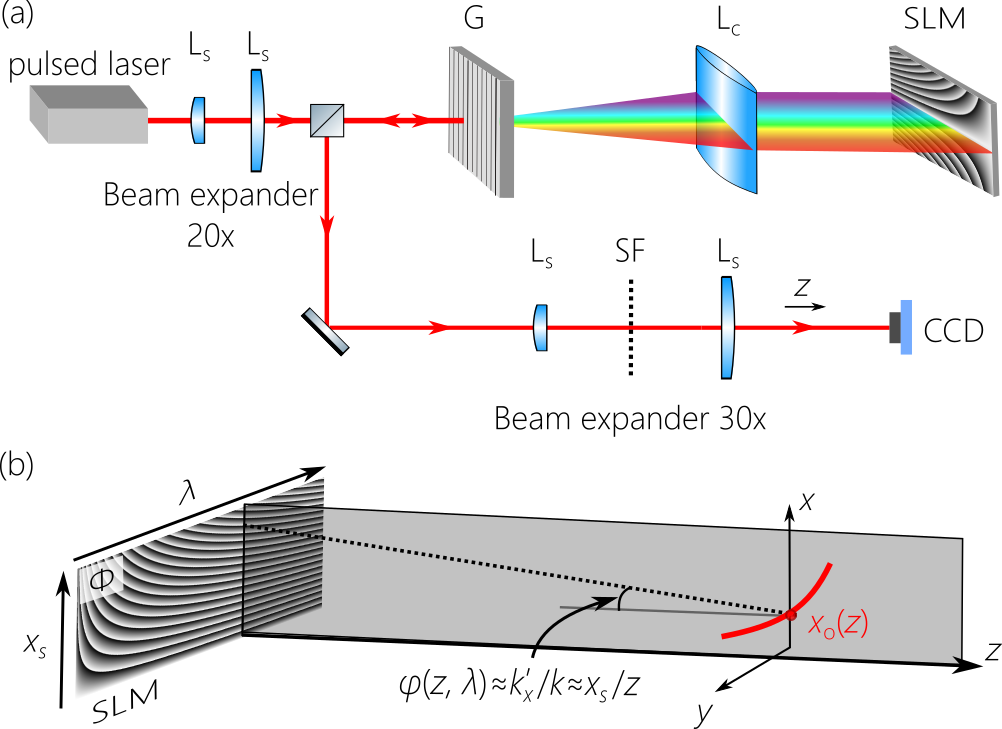}
\caption{(a) Schematic of the setup for synthesizing STWPs. G: Grating; L$_{\mathrm{c}}$: cylindrical lens; L$_{\mathrm{s}}$: spherical lens; SLM: spatial light modulator; SF: spatial filter. (b) Conceptual scheme for relating the SLM phase at the source to the local trajectory of the STWP downstream.}
\label{Fig:Setup}
\end{figure}

\subsection{Experimental synthesis methodology}

Locally bending STWPs are produced using the universal angular-dispersion synthesizer described in Refs.~\cite{Hall24JOSAA,Romer25JO}; see Fig.~\ref{Fig:Setup}(a). We start with ultrashort pulses from a mode-locked laser (Spark Laser, Alcor), which provides pulses of width 100~fs, bandwidth $\Delta\lambda\approx25$~nm, and central wavelength $\approx1064$~nm at 80~MHz repetition rate. The initial beam width of $\approx2$~mm is expanded to $\approx40$~mm before impinging on a diffraction grating (1200~lines/mm, $50\times50$~mm$^{2}$). The -1~diffraction order is selected, and then collimated via a cylindrical lens (focal length $f=500$~mm, 25-mm aperture) to spatially resolve the spectrum. A reflective, phase-only SLM (Meadowlark, E19X12) placed at the focal plane of the lens imparts a 2D phase distribution $\Phi(x_{\mathrm{s}},\lambda)$ to the spectrally resolved wave front; here $x_{\mathrm{s}}$ is the coordinate orthogonal to the direction along which the spectrum is spatially spread [Fig.~\ref{Fig:Setup}(b)]. The wave front retro-reflected back through the cylindrical lens is spectrally recombined at the grating, thereby constituting the STWP. 

Modifying $\Phi(x_{\mathrm{s}},\lambda)$ is responsible for tuning the propagation characteristics of the STWP. Each SLM column modifies the spatial phase distribution for one wavelength; we consider a single such wavelength in Fig.~\ref{Fig:Setup}(b). When the SLM phase $\Phi(x_{\mathrm{s}},\lambda)=k_{x}(\lambda)x_{\mathrm{s}}$ is linear in $x_{\mathrm{s}}$, with $k_{x}(\lambda)$ given by Eq.~\ref{eq:Parabola}, then a \textit{single} spatial frequency $k_{x}(\lambda)$ is associated with each $\lambda$, which is necessary for a propagation-invariant STWP \cite{Kondakci19OL,Yessenov19OE}. When $\Phi(x_{\mathrm{s}},\lambda)$ is \textit{not} linear in $x_{\mathrm{s}}$, then a finite spatial spectrum is associated with each $\lambda$. 

\subsection{Algorithm for designing the spatiotemporal spectral phase to realize a locally bending STWP}

Designing the SLM phase $\Phi(x_{\mathrm{s}},\lambda)$ to produce a locally bending STWP is achieved via the following algorithmic procedure:
\begin{enumerate}
\item Select the baseline STWP parameters, the spectral tilt angle $\theta$ and the temporal bandwidth $\Delta\lambda$, and calculate the transverse and axial wave numbers $k_{x}(\lambda)$ and $k_{z}(\lambda)$, respectively, associated with each wavelength $\lambda$.
\item Specify the curved beam trajectory $x_{\mathrm{o}}(z)$ for the locally bending STWP, from which we extract a local bending angle $\varphi(z)$, where $\tan\varphi(z)=x_{\mathrm{o}}(z)/z$.
\item For each wavelength $\lambda$, determine the corresponding spatial frequency $k_{x}'(z,\lambda)=k_{z}(\lambda)\sin\varphi(z)+k_{x}(z)\cos\varphi(z)$, which is now $z$-dependent.
\item We convert the $z$-dependence of $k_{x}'$ to an $x_{\mathrm{s}}$-dependence via $z\approx\tfrac{k_{x}'}{k}x_{\mathrm{s}}$ [Fig.~\ref{Fig:Setup}(b)]: $k_{x}'(z,\lambda)\rightarrow k_{x}'(x_{\mathrm{s}},\lambda)$.
\item For each wavelength $\lambda$, corresponding to a single SLM column, the phase takes the form $\Phi(x_{\mathrm{s}},\lambda)=k_{x}'(x_{\mathrm{s}},\lambda)x_{\mathrm{s}}$.
\end{enumerate}
We interpolate the SLM phase along $x_{\mathrm{s}}$. This algorithm produces the SLM phase $\Phi(x_{\mathrm{s}},\lambda)$ to first order, and it is expected that employing machine-learning will significantly improve the design of the phase distribution.

\begin{figure}[t!]
\centering
\includegraphics[width=8.6cm]{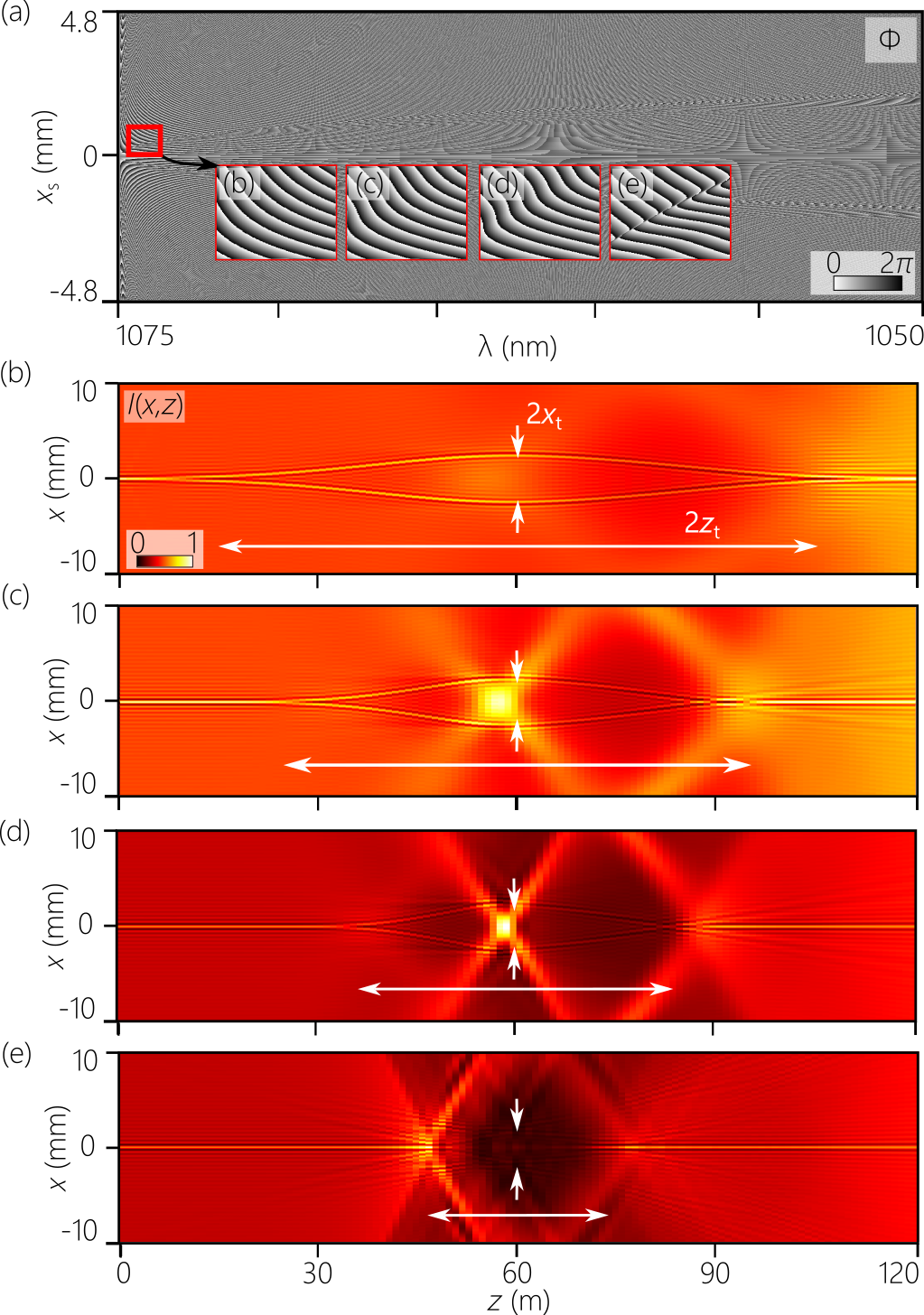}
\caption{(a) SLM phase distribution to produce a locally bending STWP of the form $x_{\mathrm{o}}(z)=\pm x_{\mathrm{t}}\exp\{-(z-z_{1})^{2}/(2z_{\mathrm{t}}^{2})\}$, with $\theta = 45.001^{\circ}$, $x_{\mathrm{t}}=2.5$~mm, and $z_{1}=60$~m. The insets correspond to the portion of the SLM phase enclosed in the red square for the four beam configurations in (b-e). (b) The calculated intensity distribution $I(x,z)$ of the locally bending STWP for $z_{\mathrm{t}}=60$~m, (c) $z_{\mathrm{t}}=20$~m, (d) $z_{\mathrm{t}}=15$~m, and (e) $z_{\mathrm{t}}=5$~m.}
\label{Fig:LimitSLM}
\end{figure}

\subsection{Emergence of an intensity null accompanying local beam bending}

Using this heuristic procedure, we calculate the SLM phase distribution $\Phi(x_{\mathrm{s}},\lambda)$ corresponding to a locally bending STWP whose peak follows the dual-branched trajectory $x_{\mathrm{o}}=\pm x_{\mathrm{t}}\exp\{-(z-z_{1})^{2}/(2z_{\mathrm{t}}^2)\}$, where $z_{1}$ is the axial plane where the obstacle is located, and $2x_{\mathrm{t}}$ and $z_{\mathrm{t}}$ are the transverse and longitudinal extents of the null produced by the locally bending beam. It is expected that the object to be avoided is localized within this null. We calculate the SLM phase $\Phi$ using the following parameters for the STWP: $\Delta\lambda=25$~nm, $\theta=45.001^{\circ}$, $z_{1}=60$~m, $x_{\mathrm{t}}=2.5$~mm, and $z_{\mathrm{t}}=60$~m [Fig.~\ref{Fig:LimitSLM}(a)]. The curvature of the locally bending beam is very mild, and its peak intensity follows very closely the target trajectory, and the constant intensity pedestal persists within the local bend [Fig.~\ref{Fig:LimitSLM}(b)]. Examining the associated SLM phase [Fig.~\ref{Fig:LimitSLM}(a)], we find that it is smooth everywhere [Fig.~\ref{Fig:LimitSLM}(a), inset], so that the spatially discretized phase implemented by the SLM can reproduce the desired distribution with high fidelity.

\begin{figure}[t!]
\centering
\includegraphics[width=8.8cm]{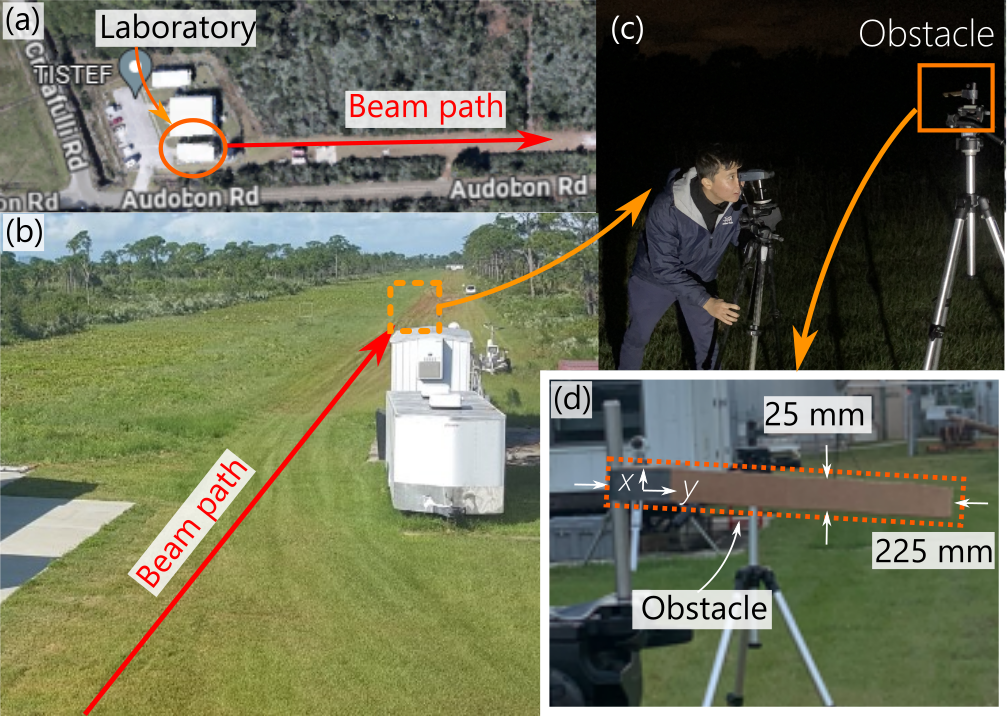}
\caption{(a) Aerial photograph of the TISTEF laser range highlighting the laboratory and the 1-km-long open-air free path (identified with a red arrow). (b) View down the beam path along the laser range from the laboratory. (c) The measurement technique at the target and the CCD camera. (d) The obstacle is attached to an optical pole on a tripod. }
\label{Fig:TISTEF}
\end{figure}

Keeping all the parameters in for the locally bending STWP in Fig.~\ref{Fig:LimitSLM}(b) fixed while reducing the value of $z_{\mathrm{t}}$ has profound impact besides the expected reduction in the axial extent of the null. First, we note that the peak intensity at the center of the local bend drops [Fig.~\ref{Fig:LimitSLM}(c-e)]. When we reach $z_{\mathrm{t}}=5$~m, the constant pedestal intensity is almost eliminated, and instead we have a null at the obstacle locations. Examining the accompanying changes in the SLM phase can help explain why this takes place. As seen in the insets to Fig.~\ref{Fig:LimitSLM}(a), the SLM phase varies more rapidly as $z_{\mathrm{t}}$ is reduced. These rapid variations cannot be captured by a realistic SLM having a finite number of pixels, and the intensity distribution produced therefore falls short of the ideal one. Specifically, the flat intensity cannot be maintained across this sharp bend, giving way to a void in the field instead. The intensity distribution away from the sharp turn taken by the beam is nevertheless produced with high fidelity.

\section{Experimental results}

\subsection{Location and setup}

\begin{figure*}[t!]
\centering
\includegraphics[width=17.6cm]{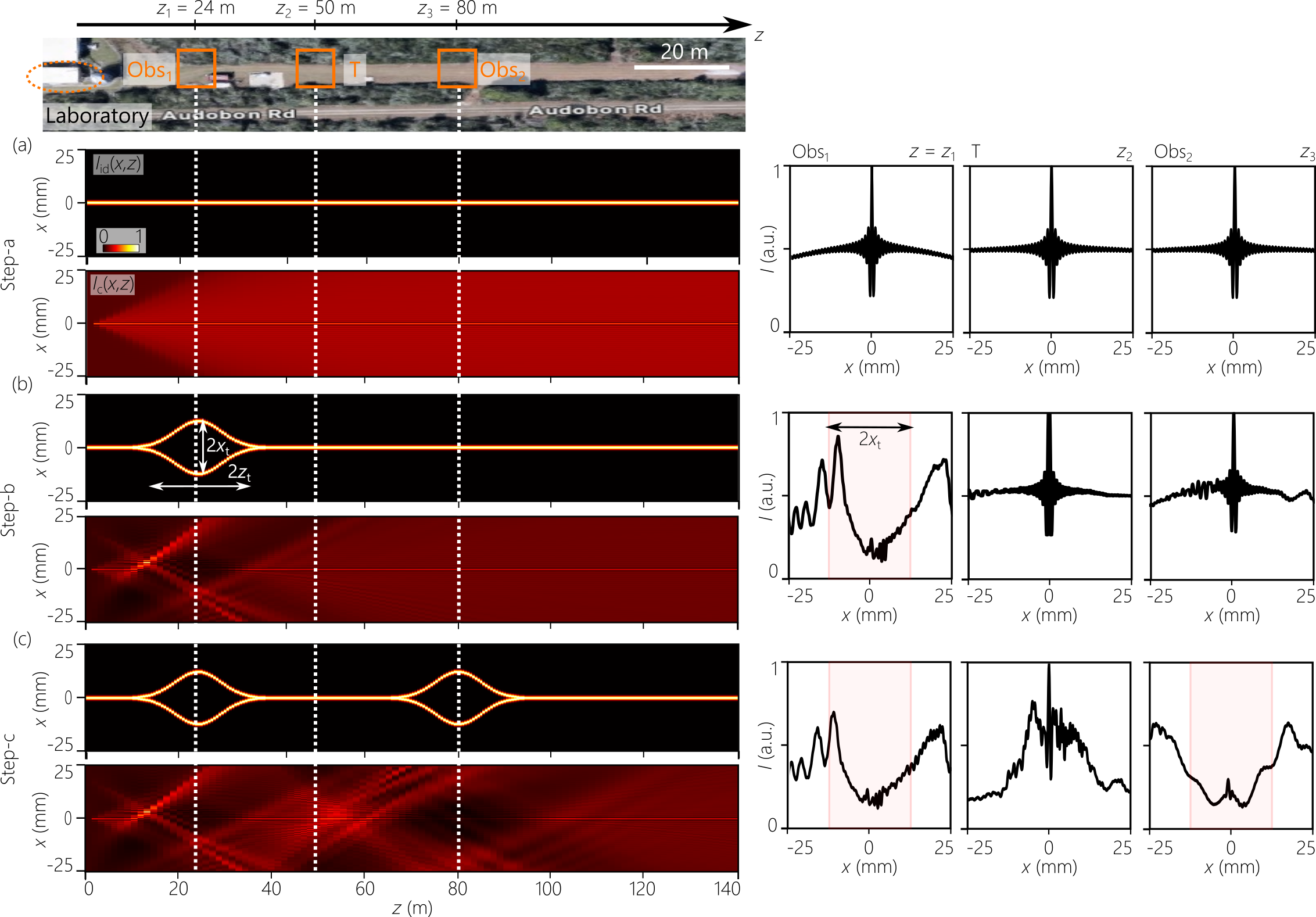}
\caption{TISTEF Scenario~I for multi-obstacle avoidance. Top image is an aerial photograph of the TISTEF laser range identifying the locations of Obs$_{1}$, T, and Obs$_{2}$ at $z_{1}=24$~m, $z_{2}=50$~m, and $z_{3}=80$~m, respectively, from the laboratory. (a) The ideal intensity $I_{\mathrm{id}}(x,z)$, the calculated intensity $I_{\mathrm{c}}(x,z)$, and sections through both at $z_{1}$, $z_{2}$, and $z_{3}$ in Step-a. (b) Same as (a) for Step-b where Obs$_{1}$ is avoided. (c) Same as (a) for Step-c where both Obs$_{1}$ and Obs$_{2}$ are avoided.}
\label{Fig:ScenarioII_Theory}
\end{figure*}

We carry out the LoS multi-target-avoidance experiments at the Townes Institute Science and Technology Experimentation Facility (TISTEF), situated on Florida's space coast [Fig.~\ref{Fig:TISTEF}(a)]. The STWP is produced by the setup depicted in Fig.~\ref{Fig:Setup}(a), which is housed in an on-site laboratory  [Fig.~\ref{Fig:TISTEF}(a)]. The STWP is magnified $\approx30\times$ via a 50-mm-diameter positive lens with $f=100$~mm followed by a 317.5-mm-diameter positive mirror with $f=3.175$~m (Edmund Optics 35-730-000). The resulting baseline propagation-invariant STWP corresponds to a spectral tilt angle $\theta=45.001^{\circ}$, a bandwidth $\Delta\lambda\approx25$~nm, $\lambda_{\mathrm{o}}\approx1064$~nm, an aperture diameter $\approx300$~mm, and a central feature of width $\Delta x\approx3$~mm, expected to have a diffraction-free propagation length of $\approx1$~km \cite{Hall25OE1km}.

The STWP is directed through an open window in the laboratory wall and then down the range where the beam path is above grassy land that runs parallel to a waterway [Fig.~\ref{Fig:TISTEF}(b)]. A computer-controlled CCD camera (TheImagingSource, DMK 33UX178) is mounted on a 3-axis tripod and transported down the range. The CCD camera records the intensity of the beam on the targets and obstacles over time at a fixed exposure [Fig.~\ref{Fig:TISTEF}(c)].

We carried out measurements in two different obstacle-avoidance configurations that we refer to as Scenario~I and Scenario~II, which involve selectively avoiding one or two LoS obstacles while maintaining incidence on a LoS target. The obstacles take the form of cardboard cutouts of dimensions $25\times225$~mm$^{2}$ attached to optical posts via adhesives and held on 3-axis tri-pods for accurate positioning in the beam path [Fig.~\ref{Fig:TISTEF}(d)]. The 25-mm-side is aligned along the $x$-axis to block the central portion of the STWP, while the 225-mm-side corresponds to the $y$-axis along which the STWP intensity profile is uniform. In each scenario, three steps are involved, each of which is associated with a different SLM phase distribution $\Phi$ that produces the spatiotemporal spectrum required to yield the beam configuration. The first step is the same for both, where a propagation-invariant STWP is sent down the range. This step allows us to align the obstacles and target in the LoS of the source. 

\subsection{Multi-target-avoidance: Scenario I}

The obstacles and target in Scenario~I are placed at the following axial locations: the first obstacle Obs$_{1}$ at $z_{1}=20$~m, the target T at $z_{2}=50$~m, and the second obstacle Obs$_{2}$ at $z_{3}=80$~m  [Fig.~\ref{Fig:ScenarioII_Theory}, top]. The goal here is to maintain incidence on T while avoiding the preceding and succeeding obstacles Obs$_{1}$ and Obs$_{2}$, respectively. The obstacles and target here all take the form of $25\times225$~mm$^{2}$ cardboard cutouts. The goal in this scenario is to maintain incidence at a fixed intensity level at T while successively avoiding Obs$_{1}$ and then Obs$_{2}$.

\begin{figure}[t!]
\centering
\includegraphics[width=8.6cm]{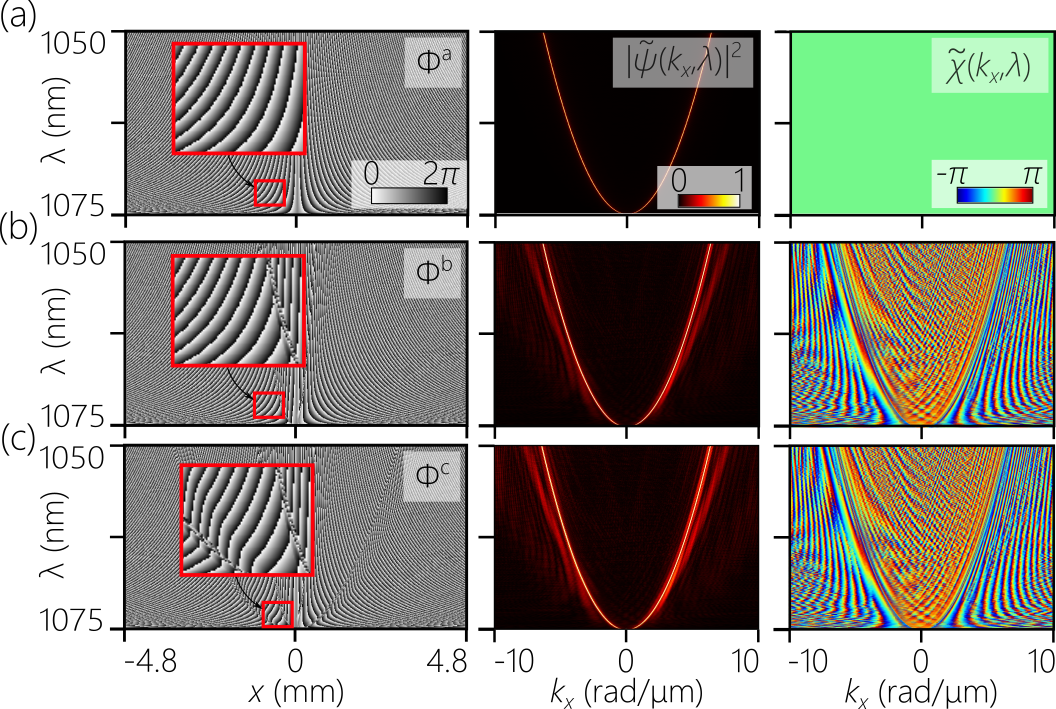}
\caption{(a) SLM phase distribution $\Phi$, and the spatiotemporal spectral intensity $|\widetilde{\psi}(k_{x},\lambda)|^{2}$ and phase $\widetilde{\chi}(k_{x},\lambda)$ for Step-a in Scenario-I [Fig.~\ref{Fig:ScenarioII_Theory}(a)], (b) for Step-b [Fig.~\ref{Fig:ScenarioII_Theory}(b)], and (c) for Step-c [Fig.~\ref{Fig:ScenarioII_Theory}(c)].}
\label{Fig:ScenarioII_Theory2}
\end{figure}

\begin{figure*}[t!]
\centering
\includegraphics[width=16cm]{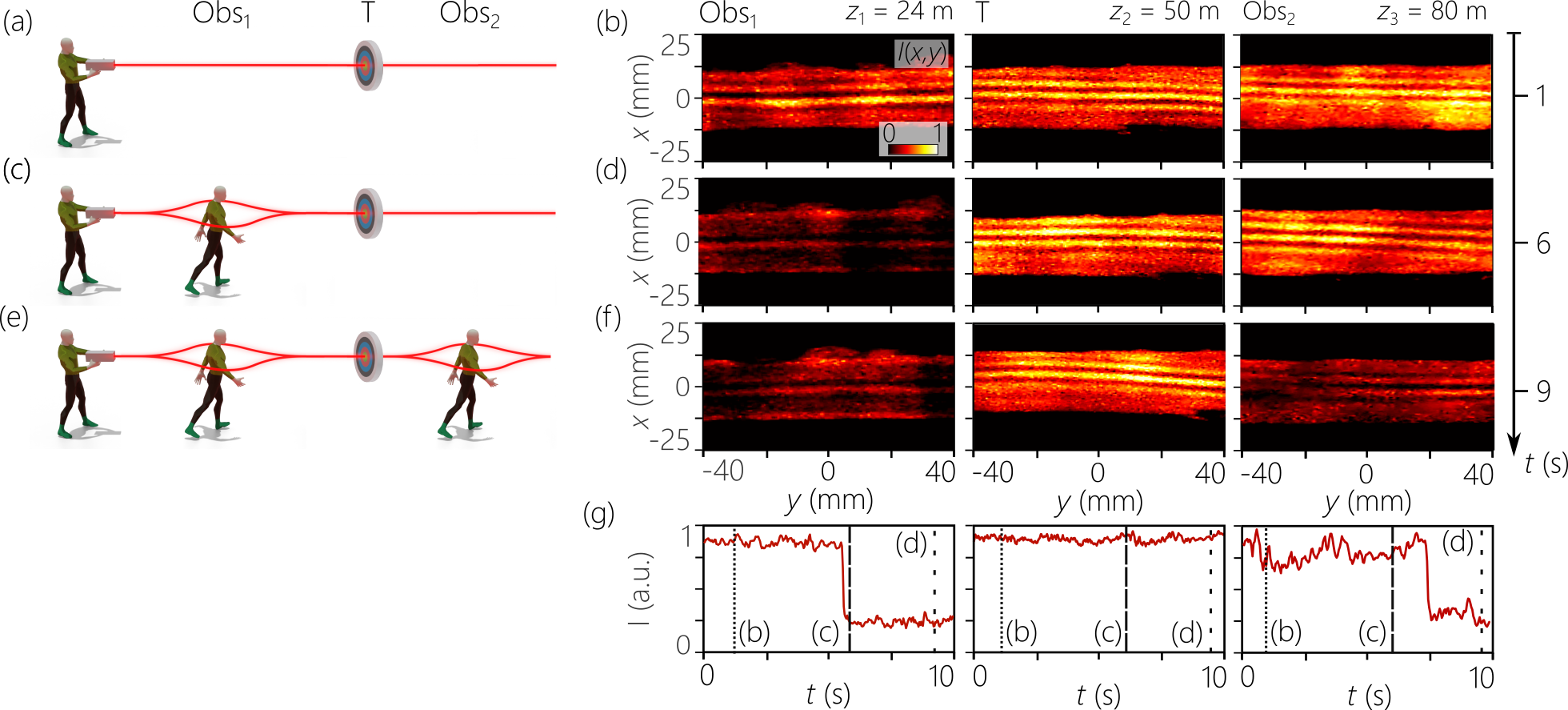}
\caption{(a) Illustration of the configuration for Step-a in Scenario-I. We depict the target as a bullseye, and the obstacles to be avoided are depicted as human forms. (b) Measured transverse intensity distribution $I(x,y,z)$ at $z_{1}$, $z_{2}$, and $z_{3}$. (c,d) Same as (a,b) for Step-b. (e,f) Same as (a,b) for Step-c. (g) Measured power levels $I(z)$ at $z_{1}$, $z_{2}$, and $z_{3}$ integrated over the transverse plane recorded over time. We highlight the instances in time corresponding to the measurements in (b,d,f).}
\label{Fig:ScenarioII_Exp}
\end{figure*}

This scenario is carried out in three successive steps: (1) in Step-a we launch a propagation-invariant STWP down the range to $z_{3}=80$~m as a baseline and to align the three targets in the LoS [Fig.~\ref{Fig:ScenarioII_Theory}(a)]; (2) in Step-b we introduce a null at $z_{1}=20$~m of transverse width $\approx25$~mm and axial extent $\approx5$~m, to avoid Obs$_{1}$ alone while maintaining incidence on both T and Obs$_{2}$ [Fig.~\ref{Fig:ScenarioII_Theory}(b)]; and (3) in Step-c we introduce a second null at $z_{3}=80$~m (having the same transverse and axial widths as that introduced at $z_{1}$ in Step-b) in addition to the null at $z_{1}$ to avoid both Obs$_{1}$ and Obs$_{2}$ while maintaining incidence on T alone [Fig.~\ref{Fig:ScenarioII_Theory}(c)]. For each of these steps, we plot the following: (1) the ideal trajectory of the beam peak $I_{\mathrm{id}}(x,z)$; (2) the calculated time-averaged intensity profile of the STWP $I_{\mathrm{c}}(x,z)$, incorporating the physical parameters of the SLM (active area and pixel size); and (3) sections through $I_{\mathrm{c}}(x,z)$ taken at $z_{1}$, $z_{2}$, and $z_{3}$.

In Step-a, the SLM phase $\Phi^{\mathrm{a}}(x_{\mathrm{s}},\lambda)$ [Fig.~\ref{Fig:ScenarioII_Theory2}(a)] corresponds to the parabolic spatiotemporal spectrum in Eq.~\ref{eq:Parabola} is implemented to produce the baseline, propagation-invariant STWP. Here the spatiotemporal spectral intensity $|\widetilde{\psi}(k_{x},\lambda)|^{2}$ is parabolic and the associated spectral phase $\widetilde{\chi}(k_{x},\lambda)$ is flat. The transverse intensity profile along $x$ takes the form of a central spatial feature of width $\Delta x\approx3$~mm atop a broad pedestal extending for $\approx300$~mm (the aperture diameter). The ideal axial intensity distribution has a peak traveling rectilinearly along the $z$-axis. The actual STWP beam has that form except for the presence of a pedestal, as seen in the sections through $z_{1}$, $z_{2}$, and $z_{3}$ [Fig.~\ref{Fig:ScenarioII_Theory}(a)]. The pedestal can be reduced by increasing the spectral uncertainty $\delta\lambda$, which is the unavoidable `fuzziness' in the association between the spatial frequency $k_{x}$ and the wavelength $\lambda$ \cite{Kondakci19OL,Yessenov19OE}. Increasing $\delta\lambda$ (e.g., by reducing the width of the grating or reducing the aperture along $x$) reduces the pedestal \cite{Kondakci19OL} but concomitantly reduces the propagation distance $L_{\mathrm{max}}$ \cite{Yessenov19OE}. We have opted here to reduce $\delta\lambda$ to guarantee a larger $L_{\mathrm{max}}$, which extends here to $\sim1$~km, although the measurements for obstacle-avoidance were carried out to $\sim130$~m.

In Step-b where we introduce a null at $z_{1}$, the SLM phase $\Phi^{\mathrm{b}}(x_{\mathrm{s}},\lambda)$ is designed according to the algorithm outlined above, and the phase distribution thus deviates away from the standard SLM phase $\Phi^{\mathrm{a}}(x_{\mathrm{s}},\lambda)$ [Fig.~\ref{Fig:ScenarioII_Theory2}(b)]. The spectral intensity is no longer purely a parabola; rather, each wavelength is now associated with a finite spatial bandwidth. The spectral phase also acquires a complex structure. It should be clear from $|\widetilde{\psi}(k_{x},\lambda)|^{2}$ and $\widetilde{\chi}(k_{x},\lambda)$ in Fig.~\ref{Fig:ScenarioII_Theory2}(b), when compared to that in Fig.~\ref{Fig:ScenarioII_Theory2}(a), that synthesizing the STWP in the spatiotemporal spectral plane $(k_{x},\lambda)$ would be extremely difficult, which motivates our selection for performing the spectral modulation in the mixed plane $(x_{\mathrm{s}},\lambda)$ instead. The ideal intensity distribution corresponds to a beam traveling rectilinearly everywhere except in the vicinity of $z_{1}$ where it locally bends, thereby opening up a gap of transverse width $\approx25$~mm, as seen in the sections through $I_{\mathrm{c}}(x,z)$ at $z=z_{1}$. Note that the gap in $I_{\mathrm{c}}(x,z_{1})$ extends in amplitude below the pedestal for the reasons explained above. At other locations, $I_{\mathrm{c}}(x,z)$ in Step-b retains the on-axis peak as in Step-a.

Finally, in Step-c we make use of the SLM phase $\Phi^{\mathrm{c}}(x_{\mathrm{s}},\lambda)$ [Fig.~\ref{Fig:ScenarioII_Theory2}(c)] to introduce the second null at $z_{3}$. The spatiotemporal spectral intensity and phases have similar complex structures to those for Step-b, with minute differences that are difficult to resolve; see the insets in the SLM phases in Fig.~\ref{Fig:ScenarioII_Theory2}(a-c). The ideal intensity distribution corresponds to a beam traveling rectilinearly everywhere except in the vicinity of $z_{1}$ and $z_{3}$, where it locally bends in a similar fashion, opening up two gaps of transverse width $\approx25$~mm. This is clear in the sections through $I_{\mathrm{c}}(x,z)$ at $z=z_{1}$ and $z=z_{3}$. In both locations, the gaps in $I_{\mathrm{c}}(x,z_{1})$ and $I_{\mathrm{c}}(x,z_{3})$ extend in amplitude below the pedestal in Step-c. In Step-a through Step-c, $I_{\mathrm{c}}(x,z)$ retains the on-axis peak at $z=z_{2}$. Therefore, the beam incidence is maintained constant throughout on a target T at $z_{2}$, even when obstacles are introduced. 

We experimentally confirm these expectations in Fig.~\ref{Fig:ScenarioII_Exp} where we plot the measured intensities at $z_{1}$, $z_{2}$, and $z_{3}$ in the three steps of Scenario-I. In the Step-a configuration illustrated in Fig.~\ref{Fig:ScenarioII_Exp}(a), the propagation-invariant STWP travels down the range. As mentioned above, Step-a allows us to align the obstacles and target, all of which have 25-mm-width along $x$. Measurements of the transverse intensity profiles $I(x,y,z_{1})$, $I(x,y,z_{2})$, and $I(x,y,z_{3})$ captured by the CCD from reflected light are plotted in Fig.~\ref{Fig:ScenarioII_Exp}(b). The targets are sufficiently small (width along the $x$ dimension of $\approx25$~mm) that the STWP self-heals \cite{Kondakci18OL} after being intercepted by each target before reaching the next target. Consequently, with all three targets in place, high intensity is recorded at all three axial planes [Fig.~\ref{Fig:ScenarioII_Exp}(b)]. In the Step-b configuration illustrated in Fig.~\ref{Fig:ScenarioII_Exp}(c), we selectively avoid obs$_{1}$ at $z_{1}=20$~m. The measurements in Fig.~\ref{Fig:ScenarioII_Exp}(d) show a drop in intensity at Obs$_{1}$, while those at T and Obs$_{2}$ have not changed. The intensity drops to $\approx25\%$ its initial value across the entire area of Obs$_{1}$. Finally, in the Step-c configuration illustrated in Fig.~\ref{Fig:ScenarioII_Exp}(e), we selectively avoid both Obs$_{1}$ and Obs$_{2}$ while maintaining incidence on T. The measured intensity distributions plotted in Fig.~\ref{Fig:ScenarioII_Exp}(f) show a drop in intensity to $\approx25\%$ the initial value at Obs$_{1}$ and Obs$_{2}$ across their entire areas. The intensity at T is maintained constant throughout.

\begin{figure*}[t!]
\centering
\includegraphics[width=17.6cm]{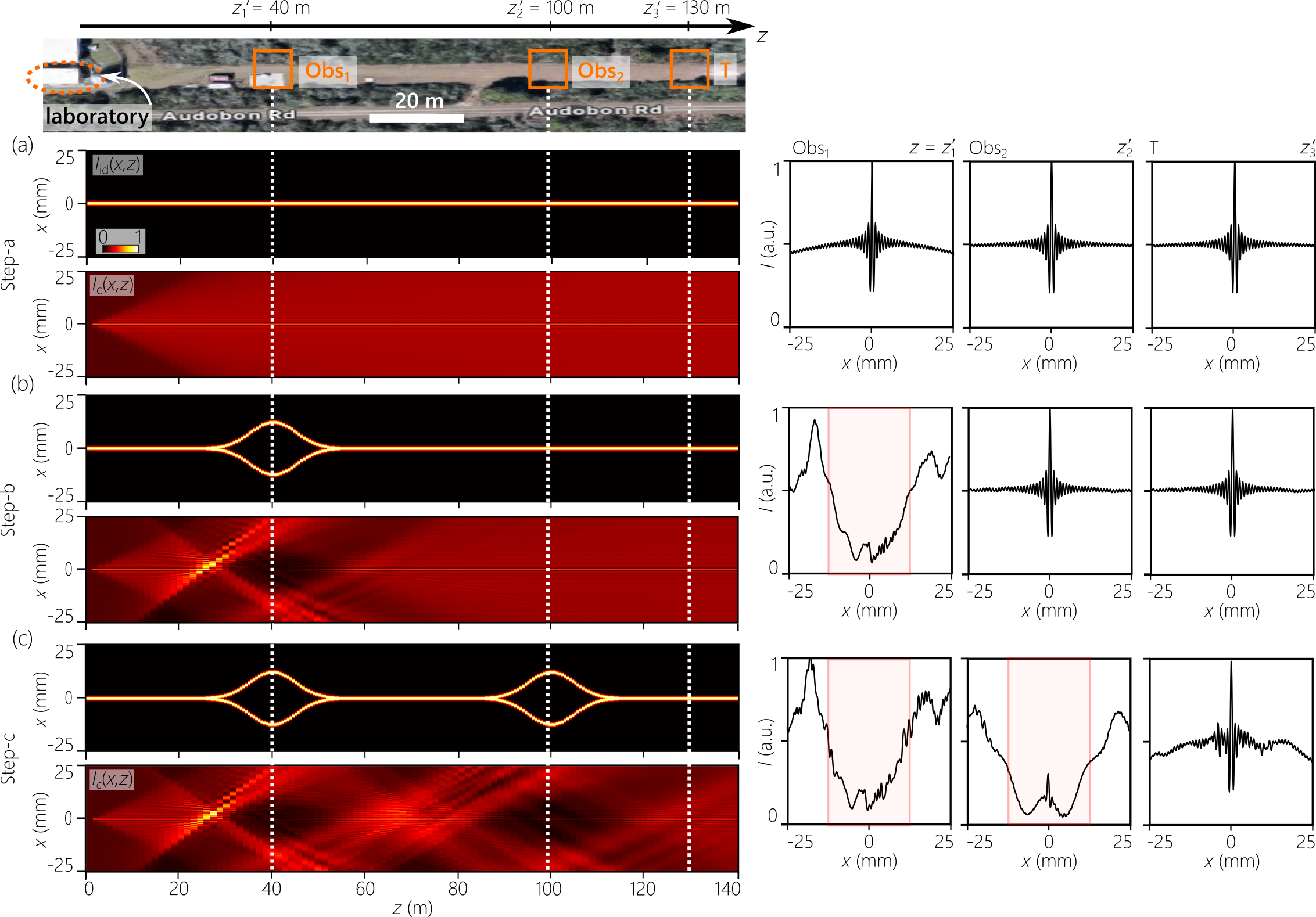}
\caption{TISTEF Scenario~II for multi-obstacle avoidance. The top aerial photograph of the TISTEF laser range identifies the locations of Obs$_{1}$, Obs$_{2}$, and T at $z'_{1}=40$~m, $z'_{2}=100$~m, and $z'_{3}=130$~m, respectively, from the laboratory. (a-c) The panels have the same structure as in Fig.~\ref{Fig:ScenarioII_Theory}(a-c): (a) Step-a, (b) Step-b where Obs$_{1}$ is avoided, and (c) Step-c where both Obs$_{1}$ and Obs$_{2}$ are avoided.}
\label{Fig:ScenarioI_Theory}
\end{figure*}

\begin{figure}[t!]
\centering
\includegraphics[width=8.6cm]{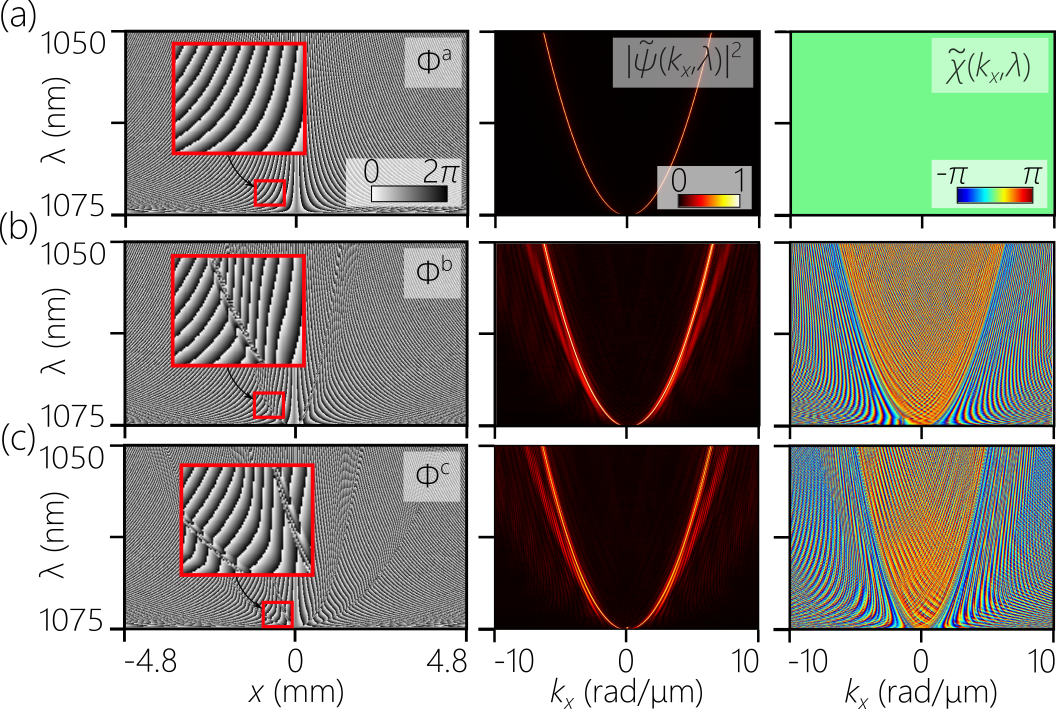}
\caption{Same as Fig.~\ref{Fig:ScenarioII_Theory2} for Scenario-II.}
\label{Fig:ScenarioI_Theory2}
\end{figure}

We plot the integrated intensity $I(z)=\iint\!dxdy\;I(x,y,z)$ at Obs$_{1}$, T, and Obs$_{2}$ in Fig.~\ref{Fig:ScenarioII_Exp}(g). At $t=0$, the phase $\Phi^{\mathrm{a}}$ [Fig.~\ref{Fig:ScenarioII_Theory2}(a)] associated with Step-a is implemented, and the power levels delivered to the three locations are all high. At $t=6$~s, the SLM phase is switched to $\Phi^{\mathrm{b}}$ [Fig.~\ref{Fig:ScenarioII_Theory2}(b)] associated with Step-b, whereupon the power level at Obs$_{1}$ drops, while these at T and Obs$_{2}$ are unaffected. This confirms that Obs$_{1}$ has been selectively avoided. At $t=9$~s, corresponding to Step-c, the SLM phase is switched to $\Phi^{\mathrm{c}}$ [Fig.~\ref{Fig:ScenarioII_Theory2}(c)], whereupon the power level at Obs$_{2}$ drops to the same level as that at Obs$_{1}$, the power level at Obs$_{1}$ remains low, and the power level at T is unaffected. This confirms that both Obs$_{1}$ and Obs$_{2}$ have been selectively avoided. The time snapshots plotted in Fig.~\ref{Fig:ScenarioII_Exp}(b,d,f) correspond to the instants highlighted in Fig.~\ref{Fig:ScenarioII_Exp}(g), which occur after switching the SLM phase. 

\subsection{Multi-target-avoidance: Scenario II}

In Scenario-II, the target and obstacles are placed at the following axial locations: the first obstacle Obs$_{1}$ at $z_{1}'=40$~m, the second obstacle Obs$_{2}$ at $z_{2}'=100$~m, and the target T at $z_{3}'=130$~m [Fig.~\ref{Fig:ScenarioI_Theory}, top]. The obstacles Obs$_{1}$ and Obs$_{2}$ take the form of cardboard cutouts of dimensions $25\times225$~mm$^{2}$, whereas the target T is a backstop that intercepts the entire STWP. The aim in Scenario-II is to maintain incidence on T at the end of the range, while selectively avoiding Obs$_{1}$ and Obs$_{2}$ that precede it.

\begin{figure*}[t!]
\centering
\includegraphics[width=16cm]{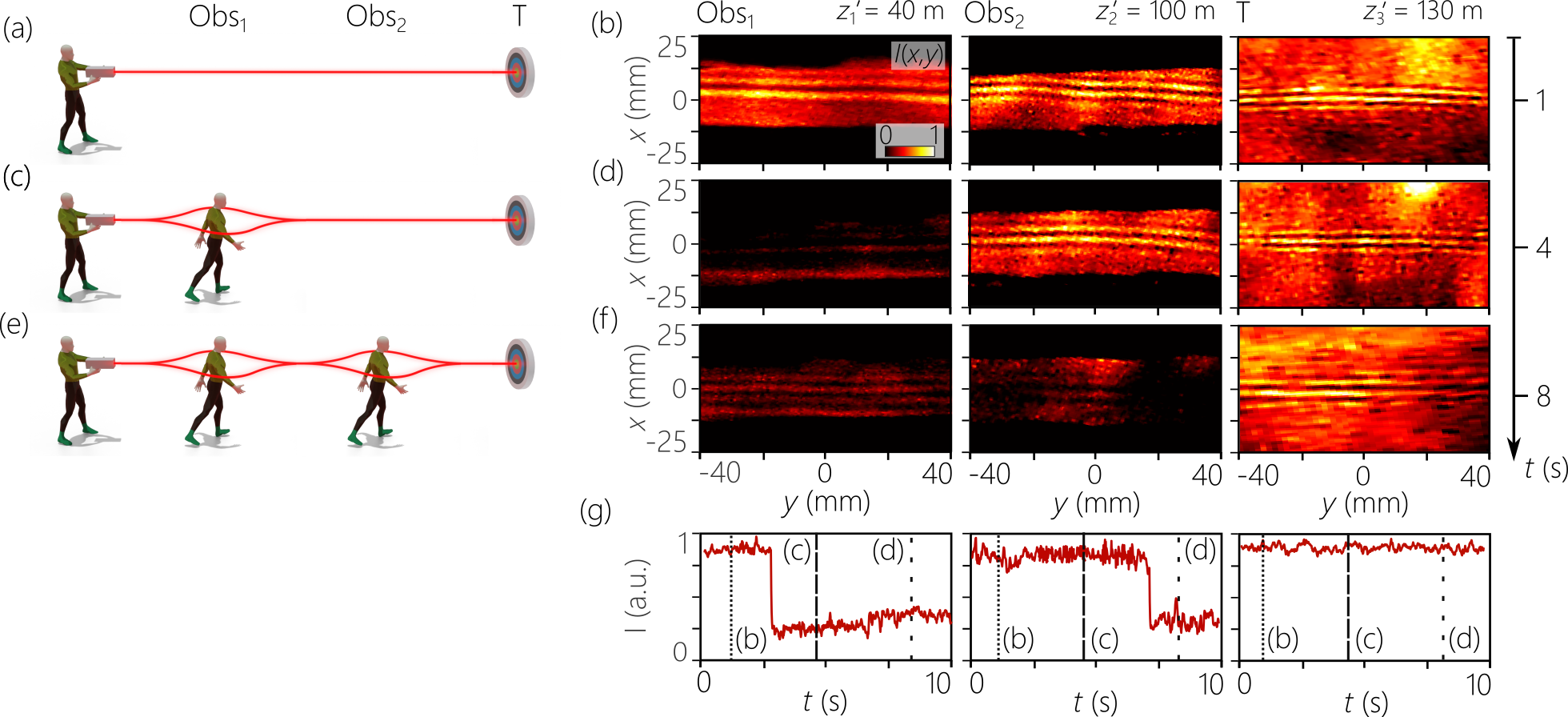}
\caption{(a) Illustration of the configuration for Step-a in Scenario-II. We depict the target as a bullseye, and the obstacles to be avoided are depicted as human forms. (b) Measured transverse intensity distribution $I(x,y,z)$ at $z_{1}'$, $z_{2}'$, and $z_{3}'$. (c,d) Same as (a,b) for Step-b. (e,f) Same as (a,b) for Step-c. (g) Measured power levels $I(z)$ at $z_{1}'$, $z_{2}'$, and $z_{3}'$ integrated over the transverse plane recorded over time. We highlight the instances in time corresponding to the measurements in (b,d,f).}
\label{Fig:ScenarioI_Exp}
\end{figure*}

This scenario is carried out in three steps: (1) in Step-a we launch a propagation-invariant STWP down the range to $z_{3}'=130$~m [Fig.~\ref{Fig:ScenarioI_Theory}(a)], which is identical to Step-a in Scenario-I except for the longer propagation distance monitored; (2) in Step-b we introduce a null of transverse width $\approx25$~mm at $z_{1}'=40$~m to avoid Obs$_{1}$ while maintaining incidence at the locations of Obs$_{2}$ and T [Fig.~\ref{Fig:ScenarioI_Theory}(b)]; and (3) in Step-c we introduce a second null (with identical parameters of the null introduced at $z_{1}'$ in Step-b) at $z_{2}'=100$~m to avoid both Obs$_{1}$ and Obs$_{2}$ while maintaining incidence on T alone [Fig.~\ref{Fig:ScenarioI_Theory}(c)]. For each step, we plot in Fig.~\ref{Fig:ScenarioI_Theory} the same quantities provided in Fig.~\ref{Fig:ScenarioII_Theory}: (1) the ideal beam trajectory $I_{\mathrm{id}}(x,z)$; (2) the calculated time-averaged intensity of the STWP $I_{\mathrm{c}}(x,z)$; and (3) sections through both of these intensity distributions at $z_{1}'$, $z_{2}'$, and $z_{3}'$.

In Step-a, a propagation-invariant STWP is directed down the range with the peak along the $z$-axis similarly to Step-a in Scenario-I except for the longer propagation distance [Fig.~\ref{Fig:ScenarioI_Theory}(a)]. Consequently, the SLM phase $\Phi^{\mathrm{a}}(x_{\mathrm{s}},\lambda)$ and the produced spatiotemporal spectrum [Fig.~\ref{Fig:ScenarioI_Theory2}(a)] are the same as for Step-a in Scenario-I [Fig.~\ref{Fig:ScenarioII_Theory2}(a)].

The goal in Step-b is to selectively avoid the obstacle Obs$_{1}$. The ideal beam intensity $I_{\mathrm{id}}(x,z)$ [Fig.~\ref{Fig:ScenarioI_Theory}(b)] locally bends around Obs$_{1}$ at $z_{1}'=40$~m resulting in an on-axis null  formed to avoid Obs$_{1}$ and a drop in the intensity of the STWP profile at Obs$_{1}$ to $\approx25\%$ of the initial power. Consequently, Obs$_{1}$ placed at this location does \textit{not} intercept the STWP. Otherwise, the STWP maintains its original behavior [Fig.~\ref{Fig:ScenarioI_Theory}(a)] before and after Obs$_{1}$, and the power levels delivered to Obs$_{2}$ and T do not change. Here $2x_{\mathrm{t}}=25$~mm and $z_{\mathrm{t}}=100$~m. The calculated intensity $I_{\mathrm{c}}(x,z)$ for Step-b shows similar behavior produced by the SLM phase $\Phi_{\mathrm{b}}(x_{\mathrm{s}},\lambda)$. The sections through $I_{\mathrm{c}}(x,z)$ highlight the drop in axial intensity at $z_{1}'$. It is instructive to compare the spatiotemporal spectra for Step-a and Step-b: the one-to-one relationship between $|k_{x}|$ and $\lambda$ is again lifted so that each wavelength is associated with a finite spectral bandwidth, and the spectral phase is no longer flat.

Finally, in Step-c we aim to avoid Obs$_{1}$ at $z_{1}'=40$~m and Obs$_{2}$ at $z_{2}'=100$~m [Fig.~\ref{Fig:ScenarioI_Theory}(c)]. Here, the STWP bends locally around the obstacles at $z_{1}'$ and $z_{2}'$, resulting in two on-axis nulls formed at $z_{1}'$ and $z_{2}'$, resulting in a drop in the STWP intensity at Obs$_{1}$ and Obs$_{2}$ to $\approx25\%$ of the initial power. The power delivered to T does not change. Consequently, Obs$_{1}$ and Obs$_{2}$ placed at these locations do \textit{not} intercept the STWP. Sections through $I_{\mathrm{c}}(x,z)$ at $z_{1}'$, $z{2}'$, and $z_{3}'$ show a drop in the intensity at $z_{1}'$ and $z_{2}'$ over a transverse width of $\approx25$~mm, while the intensity level at $z_{3}'$ remains the same.

We experimentally confirm these expectations in Fig.~\ref{Fig:ScenarioI_Exp} where we plot the measured intensities $I(x,y,z)$ at the locations of Obs$_{1}$, Obs$_{2}$, and T in the three steps of Scenario-II. In Step-a [Fig.~\ref{Fig:ScenarioI_Exp}(a)], the propagation-invariant STWP is directed to T. After incidence on Obs$_{1}$ it self-heals before reaching Obs$_{2}$, after which it again self-heals before reaching T. The measurements in Fig.~\ref{Fig:ScenarioI_Exp}(b) confirm this expectation by showing high intensity profiles at all three locations. In Step-b [Fig.~\ref{Fig:ScenarioI_Exp}(c)] we selectively avoid Obs$_{1}$, and the measurements in Fig.~\ref{Fig:ScenarioI_Exp}(d) show a drop in intensity incident on Obs$_{1}$ at $z_{1}'$ to $\approx25\%$ the value in Step-a, while those at Obs$_{2}$ and T have not changed, which instead retain the same intensity levels recorded in Step-a. Finally, in Step-c [Fig.~\ref{Fig:ScenarioI_Exp}(e)], we avoid both Obs$_{1}$ and Obs$_{2}$. The measured intensity distributions plotted in Fig.~\ref{Fig:ScenarioI_Exp}(f) show a drop in intensity at both Obs$_{1}$ and Obs$_{2}$ to $\approx25\%$ the value in Step-a, while that at T maintains its value from Step-a and Step-b.

We plot in Fig.~\ref{Fig:ScenarioI_Exp}(g) the power levels at Obs$_{1}$, Obs$_{2}$, and T over time obtained by integrating the intensity distributions over the transverse plane. At $t=0$, the phase $\Phi^{\mathrm{a}}$ [Fig.~\ref{Fig:ScenarioI_Theory}(a)] associated with Step-a is implemented, and the power levels at the three targets are all high. At $t=3$~s, the SLM phase is switched to $\Phi^{\mathrm{b}}$ [Fig.~\ref{Fig:ScenarioI_Theory}(b)] associated with Step-b, whereupon the power level at Obs$_{1}$ drops, while these at Obs$_{2}$ and T are unaffected. This confirms that Obs$_{1}$ has been selectively avoided. At $t=7$~s, corresponding to Step-c, the SLM phase is switched to $\Phi^{\mathrm{c}}$ [Fig.~\ref{Fig:ScenarioI_Theory}(c)], whereupon the power level at Obs$_{2}$ drops to the same level as that at Obs$_{1}$. The power at Obs$_{1}$ is maintained unchanged from Step-b, and that at T has been maintained fixed throughout. This confirms that both Obs$_{1}$ and Obs$_{2}$ have been selectively avoided. The time snapshots plotted in Fig.~\ref{Fig:ScenarioI_Exp}(b,d,f) correspond to the instants highlighted in Fig.~\ref{Fig:ScenarioI_Exp}(g), which occur after switching the SLM phase. 

\section{Discussion}

These results highlight that LoS obstacle-avoidance -- whether for one or multiple obstacles -- can be achieved with STWPs at large distances in an open-air environment. In principle, by tracking the LoS obstacles, the SLM phase distribution $\Phi$ can be continuously updated in real time to move the location of the nulls to locally bend around them. Our work can be extended in multiple directions of further investigations. First, we intend to extend the distance over which obstacle-avoidance is realized to 1~km \cite{Hall25OE1km}, which requires establishing the limits of this approach: (1) What is the dependence of maximum obstacle size that can be avoided on the obstacle axial location? (2) What is the minimum axial separation that allows for discrimination between an obstacle and a target? Answering these questions will determine the limit on the number, size, and locations of LoS obstacles that can be avoided in any given scenario.

As discussed earlier, more work is needed in designing the SLM phase distributions. Here we sample the location of the beam peak-intensity trajectory $x_{\mathrm{o}}(z)$ periodically along $z$, from which we determine local values of phase slopes in $k_{\mathrm{s}}(k_{x},\lambda)$ according to the heuristic procedure outlined here. We employed simple linear interpolation along each SLM column, but we anticipate that machine learning guided by initial standard configurations will help improve obstacle-avoidance, specifically by reducing the intensity at the prescribed null locations and broadening the transverse width of the null. 

Another potential approach to target-avoidance is `shadow projection': by placing a beam block in the initial beam and then relaying the structured field to the plane of the obstacle \cite{Hall26corridor}. We have recently verified this strategy using STWPs and compared its performance to that of a conventional Gaussian beam. We found that when projecting a transverse shadow that avoids a target, the length of the axial shadow cast is much smaller for an STWP than for a Gaussian beam. This sub-Rayleigh-range axial shadow cast by an STWP is related to its self-healing length \cite{Kondakci18OL}. When shadow projection of an STWP is compared to the locally bending STWP demonstrated here, we can make the following conclusions: (1) while shadow-projection involves amplitude modulation (accompanied by a power drop), locally bending STWPs are produced with phase-only modulation; (2) only single-target-avoidance has been demonstrated.

Finally, avenues for exploration include combining the local bending demonstrated here with the global bending reported in \cite{Hall25OLbending}, thereby enabling an obstacle along a curved trajectory to be avoided. In addition, combining axial and transverse acceleration and axial spectral encoding has yet to be realized. Moreover, locally bending STWPs can be potentially produced exploiting spectrally incoherent light (rather than coherent pulses), as provided by SLDs \cite{Yessenov19Optica} or LEDs \cite{Yessenov19OL}. More work is needed to produce locally bending STWPs that are localized in both transverse dimensions \cite{Yessenov22NC,Yessenov25NC}. In this case, complex beam trajectories may be produced in three dimensions. This would require generalizing the conformal mapping \cite{Bryngdahl74JOSA,Hossack87JOMO} that are a key component to the synthesis of these STWPs \cite{Yessenov22NC}.

\section{Conclusion}

In conclusion, we have realized locally bending STWPs that selectively avoid LoS obstacles intervening in the beam path. Engineering the spatiotemporal spectrum of a pulsed beam through phase-only modulation enables sculpting the beam path, thereby forming on-axis nulls at prescribed axial locations. Carrying out experiments in an open-air laser-testing facility, we have realized two different measurement scenarios. In one scenario, a designated target is located between two LoS obstacles that we aim to avoid. In the second scenario, two LoS obstacles to be avoided are introduced before the target. In both cases, incidence is maintained on the target while avoiding both obstacles. These results may inspire new approaches for free-space optical and wireless communications in the presence of LoS obstacles, and the selective delivery of radiation to a target in directed energy and radiation therapies.
  \\

\noindent\textbf{Funding}
\noindent U.S. Office of Naval Research (ONR) N00014-19-1-2192 and N00014-20-1-2789. \\

\noindent\textbf{Acknowledgments}
\noindent We thank Bryan Turo for assistance in preparing the illustrations. \\

\noindent\textbf{Disclosures}
The authors declare no conflicts of interest.\\

\noindent\textbf{Data availability}
underlying the results presented in this paper are not publicly available at this time but may be obtained from the authors upon reasonable request.


\bibliography{diffraction}

\end{document}